\documentstyle[12pt]{article}
\begin{document} 
\global\parskip 6pt
\newcommand{\be}{\begin{equation}}
\newcommand{\ee}{\end{equation}}
\newcommand{\bea}{\begin{eqnarray}}
\newcommand{\eea}{\end{eqnarray}}
\newcommand{\non}{\nonumber}

\begin{titlepage}
\hfill{hep-th/0008051}
\vspace*{1cm}
\begin{center}
{\Large\bf Exact Black Hole Entropy Bound}\\[1ex]
{\Large \bf in Conformal Field Theory}\\
\vspace*{2cm}
Danny Birmingham\footnote{E-mail: dannyb@pop3.ucd.ie}\\
\vspace*{.5cm}
{\em Department of Mathematical Physics\\
University College Dublin\\
Belfield, Dublin 4, Ireland}\\
\vspace*{1cm}
Siddhartha Sen\footnote{E-mail:
sen@maths.tcd.ie}\\
\vspace*{.5cm}
{\em School of Mathematics\\
Trinity College Dublin\\
Dublin 2, Ireland}\\
\vspace{2cm}

\begin{abstract}
We show that a Rademacher expansion can be used to establish an
exact bound for the entropy of black holes within a conformal
field theory framework. This convergent expansion includes all
subleading corrections to the Bekenstein-Hawking term.
\end{abstract}

\vspace{1cm}
August 2000 
\end{center}
\end{titlepage}

The statistical derivation of the Bekenstein-Hawking entropy
formula for black holes in string theory \cite{SV} relies heavily
on the Cardy formula for the asymptotic density of states
in two-dimensional conformal field theory \cite{Cardy}.
A particularly important example is the $(2+1)$-dimensional
BTZ black hole in anti-de Sitter space \cite{BTZ}.
The asymptotic symmetry algebra in this case consists of a Virasoro
algebra with both left-moving and right-moving sectors;
this is known as the Brown-Henneaux algebra \cite{BH}.
A simple application of the Cardy formula then yields the 
Bekenstein-Hawking entropy of the BTZ black hole \cite{Strom,BSS}.
The significance of this result follows from the fact
that many black holes in string theory
have a near-horizon structure containing
the BTZ black hole. As a result, the derivation of the entropy in
these examples is based essentially on the BTZ case
\cite{Strom,BL,CL1,CL2}.

In \cite{DMMV}, it was pointed out that 
exact convergent expressions (Rademacher expansions \cite{Rade}) 
exist for the Fourier coefficients of 
$SL(2, {\mathbf Z})$ modular forms. These were then used
to investigate in detail the AdS/CFT correspondence
for type IIB string theory on $AdS_{3} \times S^{3} \times K3$.
One important example of such
a Rademacher expansion is the exact formula for the partition 
function of an integer. Using this expansion, for example, the 
usually quoted asymptotic formula of Hardy-Ramanujan \cite{HR}
can be simply derived.
In this paper, we observe that a Rademacher expansion
gives a precise way to determine the nature of all subleading corrections
to the Bekenstein-Hawking black hole entropy. In effect, this
generalizes the Cardy formula beyond the leading term.
Furthermore, due to its convergence, the expansion leads directly
to an exact entropy bound. We also point out that the logarithmic
correction to the Bekenstein-Hawking entropy obtained
recently within a conformal field theory
framework \cite{Carlip1} follows immediately from the
Rademacher expansion. This logarithmic correction term first
appeared in the quantum geometry formalism \cite{KM}.

We are interested in how the microscopic
degrees of freedom of a conformal field theory
encode information about the entropy of a macroscopic black hole.
The starting point is to consider
the modular invariant partition function of a
unitary conformal field theory defined on a two-torus, namely
\bea
Z(\tau) = \mathrm{Tr}\;e^{2 \pi i(L_{0}
- \frac{c}{24})\tau}.
\eea
Here,  $\tau$ is the modular parameter
and $c$ is the central charge of the conformal field
theory. The Fourier expansion of the partition function
takes the form
\bea
Z(\tau) = \sum_{n\geq 0}F(n) e^{2 \pi i (n-\frac{c}{24})\tau}.
\eea
The black hole entropy in this framework is given by $S = \ln F(n)$,
for large $n$. Such a procedure is applicable
to a wide variety of black hole geometries \cite{SV}, \cite{Strom}-
\cite{CL2}, \cite{Carlip2, Sol}. To study $S$, we use an
exact convergent expansion, due to Rademacher \cite{Rade},
for the Fourier coefficients of a modular form of weight $\omega$.
This is given by \cite{DMMV}
\bea
F(n) &=& 2 \pi \sum_{m -\frac{c}{24} < 0} \left(\frac{n -\frac{c}{24}}
{|m - \frac{c}{24}|}\right)^{(\omega - 1)/2}F(m) \cdot\non\\
&\cdot & \sum_{k=1}^{\infty} \frac{1}{k}\; Kl\left(n -\frac{c}{24},
m -\frac{c}{24}; k\;\right)
I_{1-\omega}\left(\frac{4 \pi}{k}\sqrt{|m -\frac{c}{24}|
(n -\frac{c}{24})}
\;\right).
\label{Rade}
\eea
Here, $I_{1-\omega}$ is the standard Bessel function and
$Kl(n,m;k)$ is a Kloosterman sum defined by \cite{DMMV},
\bea
Kl(n,m;k) = \sum_{d \in({\mathbf Z}/k{\mathbf Z})^{*}}
\exp\left[ \frac{2 \pi i}{k}(d n + d^{-1}m)\right].
\eea

We are interested in the convergent expansion of $F(n)$ for
$\omega = 0$. For large $n$, this will
correspond to the density of states in the conformal field theory
for large eigenvalues of $L_{0}$. To begin, let us set
$m=0$ and $F(0)=1$
in (\ref{Rade}). This leads to the expression
\bea
F(n) = \frac{c}{S_{0}}\;e^{S_{0}}\; \frac{\pi^{2}}{3}\sum_{k=1}^{\infty}
\frac{1}{k} Kl(n - \frac{c}{24}, -\frac{c}{24},k) \left[e^{-S_{0}}
I_{1}\left(\frac{S_{0}}{k}
\right)\right],
\label{Rade2}
\eea
where
\bea
S_{0} = 2\pi\sqrt{\frac{c}{6}(n - \frac{c}{24})}\;.
\label{S}
\eea
Note that we have extracted a factor of $e^{S_{0}}$ from
the summation over $k$. The main point to recall
is that $e^{-x} I_{1}(x) < 1$, for all $x$ \cite{AS}.
Consequently, $e^{-x}I_{1}(x/k) < 1$, for all $x$ and for all $k$;
in fact,
one can show that $e^{-x}I_{1}(x/k) < 1/k$,
for all $x$ and for all $k$.
It is also known that the Kloosterman sum is bounded by
$k^{1/2}$ \cite{DMMV,Weil}.
It then follows that we can at least bound the argument
of the summation in (\ref{Rade2}) by $1/k^{3/2}$.
Thus, we obtain the following bound on $F(n)$,
\bea
F(n) < e^{S_{0}} \frac{c}{S_{0}}\; \frac{\pi^{2}}{3}\zeta(\frac{3}{2}),
\label{rhs}
\eea
where $\zeta(3/2)$ is the Riemann zeta function.
The corresponding entropy $S = \ln F(n)$ is then bounded as follows
\bea
S < S_{0} + \ln \left[\frac{c}{S_{0}} \frac{\pi^{2}}{3}\zeta(3/2)
\right].
\label{rhs2}
\eea 
At this point, we simply observe that the logarithmic term will
yield a negative
contribution if we impose the constraint
\bea
\frac{c}{S_{0}} < \frac{3}{\pi^{2}\zeta(3/2)}.
\label{bound2}
\eea
Subject to this constraint, the convergent Rademacher expansion leads
directly to the exact entropy bound
\bea
S < S_{0}.
\label{bound}
\eea
Below, we show that this constraint is naturally satisfied
in many examples of interest. The entropy bound is exact in
the sense that
it is derived from an exact convergent Rademacher expansion.

We should remark on the restriction of the above analysis
to the $m=0$ term in (\ref{Rade}). For general $m$, the Bessel function
term will be of the form
$I_{1}(\frac{S_{0}}{k}\sqrt{\frac{|m - \frac{c}{24}|}
{\frac{c}{24}}})$.
However, we can still extract a factor
equal to the right-hand side of (\ref{rhs}) from each of
the terms in the $m$-summation.
It follows that the terms with $m \neq 0$ are
exponentially suppressed compared
to the $m=0$ term.
We then obtain a similar result to (\ref{rhs2}),
except that the argument of the
logarithmic term contains additional exponentially supressed terms.
Since we are interested in $\ln F(n)$ for large
$n$, and hence for large $S_{0}$, these additional terms do not affect the
bound (\ref{bound}).

As an explicit example, one can consider the BTZ black hole
which is parametrized
by its mass $M = (r_{+}^{2} + r_{-}^{2})/8G{\ell}^{2}$ and
angular momentum $J= r_{+}r_{-}/4G{\ell}$. Here,
$\Lambda = -1/{\ell}^{2}$ is the cosmological constant, and $r_{\pm}$
denote the location of the inner and outer horizons.
The Brown-Henneaux algebra relates the mass and angular momentum to
the Virasoro generators by
\bea
M{\ell} = L_{0} + \bar{L}_{0} - \frac{c}{12},\;\;
J = L_{0} - \bar{L}_{0}.
\eea
Here, the normalization is that $L_{0} =  \bar{L}_{0} = c/24$
corresponds to the zero mass black hole, and $L_{0} = \bar{L}_{0} = 0$
corresponds to anti-de Sitter space.
The central charge of the Virasoro algebra is 
$c = \bar{c} = 3 {\ell}/2G$.

Although we have derived the entropy bound for a single sector,
one can consider the situation for a conformal field theory
with both left-moving and right-moving sectors.
However, for convenience, let us consider the extremal
BTZ black hole,
with $M{\ell} = J$. Then, $\bar{L}_{0} = c/24$, and moreover
\bea
S_{0} = \frac{A}{4G},
\eea
where $A = 2 \pi r_{+}$ is the length of the horizon.
Note that since we are considering macroscopic black holes,
the mass is large in Planck units, i.e., $r_{+} >> {\ell}$.
Thus, we see that the constraint (\ref{bound2})
is automatically satisfied.
As a result, the total entropy is bounded by the Bekenstein-Hawking
term $A/4G$.

It is also of interest to examine the nature of the leading
terms in the Rademacher expansion. From (\ref{Rade2}), we have
\bea
F(n) = \frac{(2 \pi)^{3/2}}{12} \;c\;S_{0}^{-3/2}\;e^{S_{0}}\;\left[1-
\frac{3}{8 S_{0}} - \cdots\right].
\label{leading}
\eea
Here, we have written only the most dominant $k=1$ term. By comparison,
the terms with
$k >1$ are exponentially damped because of the $1/k$ factor
in the Bessel function.
We have also used the asymptotic expansion of the Bessel function
$I_{1}(z) = \frac{1}{\sqrt{2 \pi z}}
\;e^{z}\;\left[1 - \frac{3}{8z} 
- \cdots\right]$, for $Re(z) \rightarrow +\infty$.
The leading terms in the black hole entropy are then given by
\bea
S = S_{0} - \frac{3}{2} \ln S_{0} + \ln c + \mathrm {constant}.
\label{Cardy}
\eea
This reveals the presence of logarithmic corrections to the
Bekenstein-Hawking entropy $S_{0}$; this is the usual correction
term which arises from the power-like factor multiplying the
asymptotic density of states \cite{DMMV}. As an example, for the extremal
BTZ black hole considered above, one finds a logarithmic
correction to the entropy of the form $-3/2 \ln(A/4G)$.
In \cite{Carlip1}, the original derivation of the Cardy formula
was extended to include the first subleading correction. Indeed,
(\ref{Cardy}) is in precise agreement with the formula
derived in \cite{Carlip1}.
One interesting point to note is that the $-3/2\ln S_{0}$ term
first appeared in the quantum geometry formalism \cite{KM}.

However, the main point to stress here is that
the Rademacher expansion is an
exact convergent expression which determines all
subleading corrections. In particular, we note that it is an expansion
in which the central charge $c$ and Virasoro generator $L_{0}$
always appear in the combination $S_{0}$
defined by (\ref{S}), with one exception. There is a lone factor
of the central charge $c$ in (\ref{leading}),
which leads to the $\ln c$ term in (\ref{Cardy}).
Thus, within the conformal field theory framework,
the coefficients of the logarithmic terms are fixed.

In \cite{Carlip2, Sol}, a conformal field theory
approach to black hole entropy in arbitrary dimensions
has been suggested. By treating the horizon
as a boundary, one finds that with a suitable
choice of boundary conditions the algebra of diffeomorphisms
in the $(r-t)$-plane near the horizon
is a Virasoro algebra. The central charge and Virasoro generator
are given by
\bea
c = \frac{3A}{2 \pi G}\frac{\beta}{\kappa},\;\;
L_{0} = \frac{A}{16 \pi G}\frac{\kappa}{\beta}.
\eea
Here, $\beta$ is an arbitrary parameter, $\kappa$ is the
surface gravity, and $A$ is the horizon area.
We note the contrast with the BTZ black hole,
where the curvature scale of anti-de Sitter space leads
to a central charge $c = 3{\ell}/2G$ which is independent
of the area.
In this case, we find the leading
term reproduces the Bekenstein-Hawking entropy, i.e.,
$S_{0} = A/4G$, provided  $\beta/\kappa <<1$.
The constraint (\ref{bound2}) is then automatically
satisfied, and we again have the exact
entropy bound $S< A/4G$.

In conclusion, we have used the convergent expansion of
Rademacher to show that the Bekenstein-Hawking entropy
provides an exact bound for the entropy of black holes within
the two-dimensional conformal field theory framework.

\noindent {\large \bf Acknowledgements}\\
This work is part of a project
supported by Enterprise Ireland Basic Research Grant SC/98/741.
D.B. would like to thank the Theory Division at CERN for
hospitality, and G. Moore and E. Verlinde for valuable discussions.

\end{document}